\documentclass[twocolumn,prb,aps,floats,showpacs]{revtex4-1}
\usepackage{amsmath}

\usepackage{epsfig}
\usepackage{graphicx}
\usepackage{dcolumn}
\usepackage{bm}
\usepackage{amssymb}
\usepackage[utf8]{inputenc}

\def\gapp{\lower.35em\hbox{$\stackrel{\textstyle>}{\sim}$}}
\def\lapp{\lower.35em\hbox{$\stackrel{\textstyle<}{\sim}$}}

\begin{document}
\bibliographystyle{apsrev4-1}

\title{Power law Kohn anomalies and the excitonic transition in graphene}
\author{F. de Juan}
\author{H.A. Fertig}
\affiliation{Department of Physics, Indiana University, Bloomington, IN 47405, USA}

\date{\today}
\begin{abstract}
Dirac electrons in graphene in the presence of Coulomb interactions of strength $\beta$ have been
shown to display power law behavior with $\beta$ dependent exponents in certain correlation
functions, which we call the mass susceptibilities of the system. In this work, we first
discuss how this phenomenon is intimately related to the excitonic insulator transition, showing the
explicit relation between the gap equation and response function approaches to this problem. We then
provide a general computation of these mass susceptibilities in the ladder approximation, and
present an analytical computation of the static exponent within a simplified kernel model, obtaining
$\eta_0 = \sqrt{1-\beta/\beta_c}$ . Finally we emphasize that the behavior of these susceptibilities
provides new experimental signatures of interactions, such as power law Kohn anomalies in the
dispersion of several phonons, which could potentially be used as a measurement of $\beta$. 

\end{abstract}

\maketitle

\section{Introduction}

One of the most intriguing aspects of graphene \cite{CGP09} continues to
be the physics of the electron-electron interaction and its effects on the electronic
properties. This very active field of research \cite{KUP10} is now of great relevance as
current experiments are displaying unambiguous signatures of interactions \cite{RUY10,EGM11}. 

What makes this problem so interesting, among other reasons, is the fact that graphene, when
undoped, is a rather special material with respect to the Coulomb interaction. Because the
low-energy quasiparticles of this system are massless Dirac fermions, with vanishing density of
states at the Fermi level, the $1/r$ Coulomb interaction remains long ranged even in the presence of
screening. This makes this system very different from materials with a finite Fermi surface, and
gives rise to very unusual many-body effects. This behavior stems from the fact that the
Hamiltonian of Coulomb interacting Dirac fermions does not contain an intrinsic energy scale. The
kinetic and interaction terms scale in the same way, and the Hamiltonian is in fact scale
invariant. In the RG sense, the Coulomb interaction is marginal in this system \cite{GGV94}. The
strength of the Coulomb interaction, the fine structure constant of graphene, may be expressed as a
dimensionless number $\beta = e^2/(\epsilon v_F)$, where $v_F$ is the Fermi velocity of the Dirac
quasiparticles, $e$ their charge, and $\epsilon$ the dielectric constant of the substrate. While a
naive estimate gives an upper limit $\beta \approx 2$ for $\epsilon=1$, implying that Coulomb
interactions should be relatively important, experimentally their strength is still a matter of
debate \cite{RUY10,EGM11}. 

The presence of this interaction has many non-trivial consequences. From a weak coupling
perspective, the first to be predicted was the logarithmic renormalization of the Fermi
velocity \cite{GGV94}, which has been recently measured \cite{EGM11}. Other observables have been
shown to be affected by a similar renormalization \cite{SS07,KUP10}. An electronic inverse
lifetime that is linear instead of quadratic with energy \cite{GGV99} is also a characteristic
feature of this system. 

Another interesting phenomenon induced by interactions is the possibility of
a phase transition to a broken symmetry state in the strong coupling regime. For a strong enough
value of the coupling $\beta$, low energy electrons and holes may bind into excitons, opening a gap
in the system. The broken symmetry state is thus known as the excitonic insulator, and the
characteristics of this transition have been studied throughly
\cite{K01,GGMS03,KS06,K09,GGG10,SSG10,ZLH10,DL09,AHS10}. The critical coupling is thought to be of
order $\beta \approx 1$ but its precise value is still under discussion
\cite{K09,GGG10,SSG10,ZLH10,DL09,AHS10}, and experimentally an intrinsic gap in graphene has not
been observed \cite{EGM11}. This excitonic transition can also be considered as the
many-body counterpart of the supercritical screening of an external Coulomb potential by Dirac
fermions \cite{GGG09,WFM10}.  

A more recent prediction of an interaction induced phenomenon, also rooted in the scale invariance
of this system, is the presence of power law behavior with interaction dependent exponents in
certain correlation functions \cite{WFM10,WFM11,G10,GMP10,GGG10}, as if the system were in a
critical phase. These correlators can be considered the mass-mass response functions (or mass
susceptibilities) of the system. In analogy to the charge-charge response function, which measures
the charge expectation value as a response to an external perturbation coupling to the charge,
this susceptibility is the analog quantity built with mass vertices (i.e. the
matrices that gap the Dirac spectrum) instead of charge ones. The experimental detection of these
correlations would represent a new signature of interaction effects beyond those already discussed,
and would be highly desirable. In addition, the singularities of the mass susceptibilities are
directly related to instabilities to spontaneous mass generation, and therefore their observation
will shed light on the problem of the excitonic transition. This observation is however challenging,
as there is no simple experiment to directly probe a mass (as opposed to charge or spin)
susceptibility.

The objective of this work is twofold. First we will show the explicit correspondence between the
physics of the excitonic transition and the behavior of the mass susceptibilities. We will
illustrate it in detail by comparing the transition as seen from a gap equation and a
response function perspective. Second, we will argue that there are in fact current experiments
available to observe the mass susceptibilities. We will see that there are clear
experimental signatures of their presence in the dispersion relation of certain phonons, which
acquire power law Kohn anomalies with $\beta$ dependent exponents. In addition, we will
show that the static exponent of the mass susceptibility, obtained within perturbative RG
\cite{G10,GMP10,GGG10}, or numerically \cite{WFM10,WFM11}, can in fact can be obtained analytically
from the ladder approximation and is given by $\sqrt{1-\beta/\beta_c}$. 

The paper is organized as follows. In Sec. \ref{sec2} we review the physics of the excitonic
transition and its description in terms of a gap equation. In Sec. \ref{sec3} we describe the
general mass-mass response function and compute it in different approximations. In Sec. \ref{sec4}
we discuss how the mass susceptibility may be observed in the dispersion relation of particular
phonons, and in Sec. \ref{sec5} we present our conclusions. 

\section{The excitonic transition}\label{sec2}

\subsection{The model}

We start by reviewing the excitonic insulator transition in graphene. We will
consider spinless graphene for simplicity, as spin will not play any role and can be accounted for
when necessary. In graphene, the low energy excitations around
the $K$ and $K'$ points can be modeled in terms of a Dirac Hamiltonian in two spatial dimensions
\begin{equation}\label{H_el}
H = i v_F \int d^2r \psi^{\dagger} ( \alpha_x \partial_x + \alpha_y \partial_y) \psi,
\end{equation}
with $\vec \alpha = (\tau_z \sigma_x, \sigma_y)$, where the $\sigma$ and $\tau$ matrices act 
on the sublattice and valley degrees of freedom, respectively. The Coulomb
interaction is included as
\begin{equation}
H_{int} = \frac{e^2}{2} \int d^2r d^2 r'
\frac{\psi^{\dagger}_r\psi_{r}\psi^{\dagger}_{r'}\psi_{r'}}{|\vec r- \vec r'|}.
\end{equation}
This Hamiltonian has an SU(2) valley symmetry generated by the matrices $T_n=(\tau_x
\sigma_y, \tau_y \sigma_y, \tau_z)$, in the sense that the SU(2) rotation $\psi
\rightarrow e^{i T_n \theta_n}\psi$ leaves the Hamiltonian invariant\footnote{Note
that this symmetry is only valid at low energies, as contact interactions, which are
irrelevant in the RG sense, may in general break it}.

When $\beta$ is large enough, this system has an instability towards the pairing of electrons
and holes. This pairing generates a mass for the Dirac fermions, opening a gap in the spectrum, and
so this state is known as the excitonic insulator. This instability is signaled by the development
of a finite expectation value of a mass operator $\left < \psi^{\dagger} M \psi \right>$. There are
four possible mass matrices $M$ for the Hamiltonian (\ref{H_el}), which correspond to different
microscopic mechanisms for the instability, and which break different symmetries. The first of them
is the generation of a charge imbalance between sublattices (a charge density wave), and corresponds
to a matrix $\sigma_z$. The next two are produced by a bond density wave, the Kekulé distortion
\cite{C00}, and correspond to the matrices $\tau_x \sigma_x$ and $\tau_y\sigma_x$. Finally, a
pattern of circulating currents in the unit cell gives rise to the mass $\sigma_z \tau_z$, known as
the Haldane mass.  

The Kekulé and CDW masses are time reversal invariant, and may be grouped into the three components
of a spin 1/2 vector
\begin{equation}
M_n = (\tau_x \sigma_x, \tau_y \sigma_x, \sigma_z),
\end{equation}
because they transform as such under the SU(2) valley symmetry. Therefore, when one of them is
generated, the valley symmetry is spontaneously broken to U(1), analogously to a ferromagnet. The
Haldane mass, on the other hand, is a scalar under valley symmetry but breaks time reversal. In
this work we will be concerned only with the time reversal invariant masses and valley symmetry
breaking. The structure of these masses and the symmetries they
break is analogous to the one found in QED3, where the valley symmetry is known
as chiral symmetry. In fact, the problem of chiral symmetry breaking and mass generation in QED3 in
the $1/N$ approximation \cite{P84,ANW88} has much in common with the excitonic
transition, the main difference being that the interaction in QED3 is a Lorentz invariant gauge
field (rather than an instantaneous charge-charge interaction). 

\subsection{The gap equation}

The most common approach to show the existence of the excitonic transition is by means of a
self-consistent gap equation \cite{K01,GGMS03,KS06,K09,GGG10,SSG10,ZLH10}. This is equivalent to the
Hartree-Fock or mean field approximation. We will now review the main features of this approach,
with the aim of highlighting its relation with the response function approach to be discussed in
the next section. For simplicity we set $v_F=1$ henceforth and thus the Coulomb coupling $e^2 =
\beta$. 

\begin{figure}[h]
\begin{center}
\includegraphics[width=8cm]{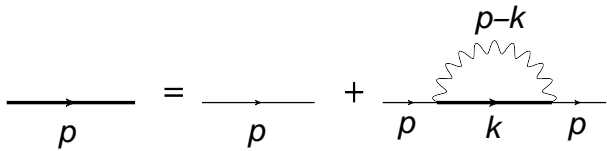}
\caption{Self consistent diagrammatic equation for the electron Green function (full line) in the
rainbow approximation}\label{rainbow}
\end{center}
\end{figure}

To derive the gap equation, the full electron propagator $G$ is expressed in terms of the
self-energy as $G^{-1} = G_0^{-1} - \Sigma$, with the bare propagator 
\begin{equation}
G_0(k)=\frac{k_0 + \vec{\alpha}\vec{k}}{k_0^2- k^2 +i\epsilon}.
\end{equation}
The self-energy is a 4x4 matrix that satisfies the Schwinger-Dyson equation
\begin{equation}
\Sigma(p) = - i \int  \frac{d^3 k}{(2\pi)^3} \tilde{\Gamma}^0(k,p) \tilde{\Pi}(p-k)
\left( G_0^{-1} - \Sigma
\right)^{-1}, 
\end{equation}
where $G,\tilde{\Pi}$ are the full electron and photon
propagators, and $\tilde{\Gamma}^0$ the full
Coulomb vertex.  This equation is so far exact, but to solve it one needs to assume approximate
forms for $\tilde{\Pi}$ and $\tilde{\Gamma}^0$. The simplest approximation is to assume a bare
Coulomb vertex $\tilde{\Gamma}^0 = 1$ and a bare photon $\tilde{\Pi} = 2\pi \beta /|\vec q|$, which
gives
\begin{equation}
\Sigma(p) = -i \beta \int \frac{d^3 k}{(2\pi)^2} \frac{1}{|\vec p-\vec k|} \left( G_0^{-1} - \Sigma
\right)^{-1}. \label{Sig_1}
\end{equation}
This equation can be seen as the resummation of the ``rainbow'' diagrams, depicted diagrammatically
in Fig. \ref{rainbow}. To solve this equation $\Sigma$ can be expanded in a basis of all sublattice
and valley matrices. We will make the further approximation that only terms proportional to mass
matrices are important (neglecting Fermi velocity and wavefunction renormalizations) i.e. we assume
that
\begin{equation}
\Sigma_{ij} = \Delta_n (M_n)_{ij},
\end{equation}
where summation over repeated indices is always implicit. Projecting into each mass channel (i.e.
taking the trace
of Eq. (\ref{Sig_1}) with $M_m$) we arrive at 
\begin{equation}
\Delta_m(p) = -i\beta \delta_{mn}  \int \frac{d^3k}{(2\pi)^2} \frac{\Delta_n(k)}{|\vec p-\vec k|}
\frac{1}{k_0^2 -
k^2 -\Delta_n(k)^2}.
\end{equation}
While the gap equations for the three masses can be studied separately, we see that they are
in fact related by symmetry, so that we can drop the index $n$. Because the interaction is $k_0$
independent in this approximation, the gap is too, and we may integrate over $k_0$ to obtain
\begin{equation}
\Delta(p) = \pi \beta \int \frac{d^2k}{(2\pi)^2} \frac{\Delta(k)}{|\vec p-\vec k|}
\frac{1}{\sqrt{ k^2+\Delta(k)^2}}. \label{gape}
\end{equation} 
This is the simplest version of the gap equation for the excitonic problem. More refined
approximations have been considered in the literature, for example at finite temperature
\cite{K01}, with the renormalization of the fermion spectrum 
\cite{K09,KS06,SSG10}, or with static \cite{GGG09,ZLH10} and dynamic \cite{GGG10} RPA (1/N)
screening for the interaction. In general, all approaches agree that there is an excitonic
transition, but critical couplings vary significantly. For our discussion we retain
the simplest form given by Eq. (\ref{gape}). 

\begin{figure}[h]
\begin{center}
\includegraphics[width=8cm]{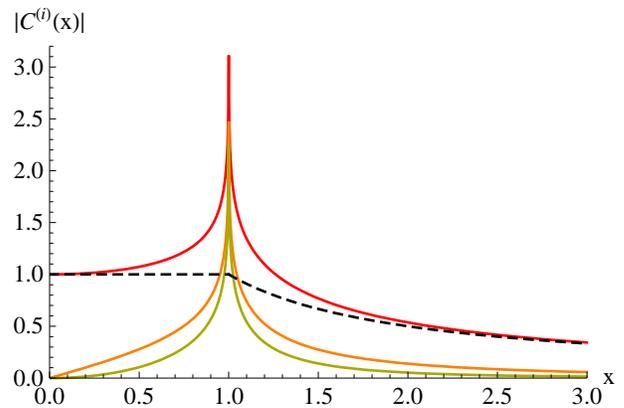}
\caption{Plots of the first three circular harmonics of the Coulomb kernel $C^{(0)}(k/p)$, Eq.
(\ref{coulombker}). The corresponding simplified kernel given in Eq. (\ref{simpkercoulomb}) is shown
for comparison
(dashed line).}\label{kernelplots}
\end{center}
\end{figure}

\begin{figure}[h]
\begin{center}
\includegraphics[width=8cm]{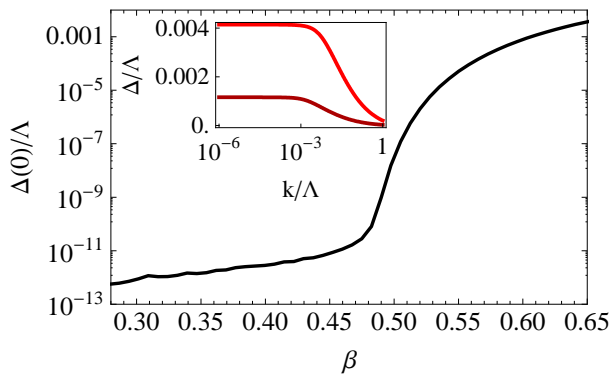}
\caption{Dependence of the gap $\Delta(0)$ on $\beta$ for $\tilde{\mu}=10^{-10}$. The gap becomes
significant within numerical precision for $\beta \gtrsim
0.48$. Note that this computation is done with the
full kernel $C^{(0)}$ and $\beta_{c}^{(0)}$ need not be greater than
$1/2$. Inset: Momentum dependence of the gap $\Delta(p)$, for
$\beta=0.61$ (lower curve) and
$\beta=0.65$ (upper curve), computed with Eq. (\ref{intgap}). 
}\label{gapplots}
\end{center}
\end{figure}

To make further progress, the standard approach is to implement an expansion of the gap in circular
harmonics, 
\begin{equation}
\Delta^{(n_p)} = \int \frac{d\theta_p}{2\pi}e^{i n_p\theta_p}\Delta(p),
\end{equation}
and keeping only the first order term, $\Delta = \Delta^{(0)}$, which is equivalent to assuming that
$\Delta$ has no angle dependence. The equation is now expressed as
\begin{equation}
\Delta(p) = \frac{\beta}{2 p} \int k dk 
\frac{\Delta(k)}{\sqrt{ k^2+\Delta(k)^2}} C^{(0)}(k/p),\label{intgap}
\end{equation} 
where the Coulomb kernel is defined as
\begin{equation}\label{coulombker}
C^{(n)}(x) = \int \frac{d \theta_k}{2\pi} \frac{e^{i n \theta_k}}{(1+x^2+ 2x \cos
\theta_k)^{1/2}},
\end{equation}
This integral equation can be solved numerically by iteration, but it is instructive to discuss
first an analytical solution that is available when a simplified version of the kernel is taken,
given by
\begin{equation}
\label{simpkercoulomb}
C^{(0)}(x) =  \theta(1-x) + \frac{1}{x}\theta(x-1),
\end{equation}
which has the correct asymptotic behavior at large and small arguments. The Coulomb kernel and its
approximate form, as well as higher order kernels that are neglected, are plotted in Fig.
\ref{kernelplots} for comparison. 
The advantage of this approximate kernel is that it allows one to re-express the integral equation
as a differential one. Taking two derivatives of Eq. (\ref{intgap}) with respect to $p$, we obtain
\begin{equation}
p^2 \frac{\partial^2 \Delta}{\partial p^2} + 2 p \frac{\partial \Delta}{\partial p} +
\frac{\beta}{2} \frac{p \Delta}{\sqrt{ p^2+\Delta^2}} = 0. \label{diffeq}
\end{equation}
The boundary conditions for this equation are obtained from the first derivatives of Eq.
(\ref{intgap}) at the endpoints of the integration region
\begin{eqnarray}
\left. p\frac{\partial \Delta}{\partial p} +\Delta \right|_{p=\Lambda}= 0, \label{ultcon}\\
\left. p^2 \frac{\partial \Delta}{\partial p} \right|_{p=\mu}= 0 \label{infcon},
\end{eqnarray}
where $\mu$ is an infrared cutoff (the inverse system size). Eq. (\ref{diffeq}) cannot be
solved analytically, but one may linearize it to obtain the branching points in $\beta$ where
non-trivial solutions become possible \cite{K01,GGG10},
\begin{equation}
p^2 \frac{\partial^2 \Delta}{\partial p^2} + 2 p \frac{\partial \Delta}{\partial p} +
\frac{\beta}{2} \Delta = 0.
\end{equation}
This equation is the well known Schroedinger equation with a
$1/r^2$ potential \cite{Landau}. Its solution is given by
\begin{equation}
\Delta = A_+ \tilde{p}^{\eta_+} + A_-\tilde{p}^{\eta_-},
\end{equation}
with $\eta_{\pm} = 1/2 (1 \pm \gamma)$, $\gamma = \sqrt{1-2\beta}$ and all quantities with a tilde
are scaled with the cutoff $\tilde{p} = p/\Lambda$. Applying the infrared condition (Eq.
\ref{infcon}) we get
\begin{equation}
\Delta = A \left[ \eta_- \tilde{p}^{\eta_+}-\eta_+ \tilde{\mu}^{\gamma}
\tilde{p}^{\eta_-} \right].
\end{equation}
The ultraviolet condition (Eq. \ref{ultcon}) is also homogeneous and so does not fix the scale of
the gap $A$, something that can only be done with the non-linear equation. The ultraviolet
condition gives is the set of branching points where non-vanishing solutions of the gap equation
become possible
\begin{equation}
\sqrt{2\beta-1} \log \tilde{\mu} = 2 \arctan \left(
\frac{\sqrt{2\beta-1}}{\beta-1}\right) +2n\pi \label{poles},
\end{equation}
which represents an infinite number of logarithmically spaced solutions $\beta_c^{(i)}$for
$\beta>1/2$. (When $\mu\rightarrow0$ all solutions collapse to $\beta_c = 1/2$, although there is
still one of
lowest energy \cite{GGMS03}). Note however that for $\beta$ greater than $\beta_c^{(0)}$, a finite
gap is always present and the linearized equation cannot be used. The rest of the solutions
thus represent higher order instabilities that would take place if the first one is set artificially
to zero. Coming back to the full integral equation, Eq. (\ref{intgap}), we can solve it by iteration
procedures, and
obtain both the gap and its dependence on $\beta$, which are plotted in Fig. \ref{gapplots}. We
observe that, for $\tilde{\mu} = 10^{-10}$, a finite gap is generated for $\beta > \beta_c^{(0)}
\approx 0.48$, which is the signature of the excitonic transition. From now on we will simply call
$\beta_c \equiv \beta_c^{(0)}$, the physically relevant critical coupling.  

\section{Response functions}\label{sec3}

A complementary approach to study the excitonic transition is through response functions. These are
defined as the expectation value of a mass operator $\psi^{\dagger} M \psi$ expanded to first order
in an external mass perturbation. They can be considered as generalized susceptibilities in the mass
channel, and their singularities represent the instabilities of the system to spontaneous mass
generation. Therefore they represent an alternative approach to study the excitonic transition.
However, the importance of the response functions goes beyond the characterization of the
transition, as they represent physical observables with non-trivial interaction signatures also for
$\beta < \beta_c$. This has also been emphasized in the context of the QED$_3$ theory
of the cuprates \cite{GHR01,FPS03,GKR03,HSF05}, where the behavior of the mass susceptibilities is
similar to what is found in graphene.

\begin{figure}[h]
\begin{center}
\includegraphics[width=8cm]{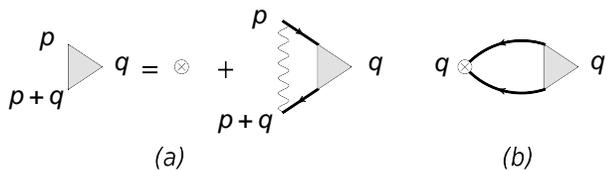}
\caption{(a) Diagrammatic equation for the three point vertex (shaded triangle) in
the ladder approximation. The cross denotes a mass vertex. (b) Response function diagram.
}\label{ladder}
\end{center}
\end{figure}

The response in the $M_n$ channel to a perturbation $M_m$ is simply the correlator 
\begin{align}\label{response}
\Pi_{nm}(q) = 2i  \int \frac{d^3p}{(2\pi)^3} \; tr \left[M_n G(p) \Gamma_m(p,p+q) G(p+q)\right],
\end{align}
which is shown diagrammatically in Fig. \ref{ladder}(b), and where $\Gamma_m$ is the full mass
vertex (note that $\Gamma_m$ is a 4x4 matrix; the sublattice/valley index is omitted for clarity)
and the factor of 2 accounts for spin.
Again, this equation is exact but to solve it we need to approximate $\Gamma$ in some way. To
guide us in the choice of approximation, we realize that the full mass vertex satisfies the
following property: if we include an external mass $m_n \psi^{\dagger}M_n \psi$ in the
Hamiltonian, then by construction it holds that
\begin{equation}
\left. \frac{\partial G(p)}{\partial m_n} \right|_{m_n = 0} = \left. G(p)\Gamma_n(p,p)G(p)
\right|_{m_n = 0}, 
\end{equation}
or equivalently in terms of the self-energy and writing the indices explicitly
\begin{equation}
\left. (M_n)_{ij} + \frac{\partial \Sigma_{ij}(p)}{\partial m_n} \right|_{m_n = 0} = \left.
(\Gamma_n)_{ij}(p,p)
\right|_{m_n = 0}. \label{sigmaprime}
\end{equation}
This identity can be proven diagrammatically by realizing the derivative acts by cutting all
possible fermion lines in the full self-energy and introducing a mass vertex $M_n$ at every cut. If
we assumed the rainbow summation for the self-energy, it can be seen that the corresponding
approximation for $\Gamma$ is the ladder approximation, depicted in Fig. \ref{ladder}(a). In this
case $\Gamma$ satisfies the self-consistent equation
\begin{equation}\label{selfc}
\Gamma_m(p,q) = M_m + i \beta \int \frac{d^3 k}{(2\pi)^2} \frac{G(k)
\Gamma_m(k,q) G(k+q)
}{|\vec p-\vec k|}.
\end{equation}
It is important to note that the propagators in this equation are the full propagators in the
rainbow approximation. These are just $G_0$ for $\beta < \beta_c$ as shown in the previous section,
but acquire a mass for $\beta > \beta_c$ and this has to be included to have a consistent
computation.
This simply represents the fact that after the phase transition, the propagators have to be
computed in the broken symmetry state. Note also that, as in what happens in the gap
equation, we anticipate by symmetry that the response function will satisfy $\Pi_{nm} = \delta_{nm}
\Pi$, so we can drop the indices in this case too. 

We will now compute the response and vertex functions, and for the sake of clarity we will do so in
two steps. To have some insight into the excitonic transition itself, it is simplest to
discuss just $\Pi(q_0=0,q=0)$. In addition the computation for $\beta>\beta_c$ presents no major
difficulty. Our main interest, however, is the full response function $\Pi(\omega,q)$, which will
compute next in what appears to be the experimentally relevant case of $\beta < \beta_c$. The
extension of $\Pi(\omega,q)$ to $\beta > \beta_c$ is also feasible but it is beyond the
scope of this work.  

\subsection{$\Pi(0,0)$ and the excitonic transition}

The computation of $\Pi(0,0)$ is simplified by the identity in Eq. (\ref{sigmaprime}). Since in the
previous section we assumed $\Sigma_{ij} = \Sigma_n (M_n)_{ij}$, Eq. (\ref{sigmaprime}) implies
that that $\Gamma_n(p,p)$ is also proportional to the corresponding mass matrix, $(\Gamma_n)_{ij}
=  (M_n)_{ij} \Gamma$. This allows us to perform the traces easily in both Eqs. (\ref{response})
and (\ref{selfc}). Plugging in $G$ as obtained in the previous section, and integrating in energy,
the equations read  
\begin{align}
\Pi(q) &= 4 \int \frac{d^2p}{(2\pi)^2} \frac{p^2}{(p^2 + \Delta^2)^{3/2}}\Gamma(p,p), 
\\
\Gamma(p,p) &= 1 + \pi \beta \int \frac{d^2k}{(2\pi)^2} \frac{k^2}{(k^2 + \Delta^2)^{3/2}}
\frac{1}{|\vec k - \vec p|}\Gamma(k,k) .
\end{align}
As for the gap equation, we now assume that $\Gamma$ has no angle dependence (i.e. we take only the
first order of the expansion in circular harmonics) we find
\begin{align}
\Pi(q) &= \frac{2}{\pi} \int dp \frac{p^3}{(p^2 + \Delta^2)^{3/2}}\Gamma(p,p),  \\
\Gamma(p,p) &= 1 + \frac{\beta}{2 p} \int dk \frac{k^3}{(k^2 + \Delta^2)^{3/2}} C^{(0)}(k/p)
\Gamma(k,k). \label{gamma00}
\end{align}
A direct computation shows that Eq. (\ref{gamma00}) can also be obtained by adding an external
mass $m_n$ to the propagator of Eq. (\ref{Sig_1}) and taking the derivative with respect to it, as
Eq. (\ref{sigmaprime}) mandates. 

We are now ready to make the equivalence between the gap equation and the response function
approaches
explicit. Consider Eq. (\ref{gamma00}) for $\Delta=0$. If we write the equation for $\Gamma$ in the
form
\begin{equation}
\int dk A(k,p) \Gamma(k,k) = 1,
\end{equation}
then
\begin{equation}
A(k,p) = \delta(k-p)- \frac{\beta}{2 p} C^{(0)}(k/p).
\end{equation}
We can obtain $\Gamma$ by inverting the operator $A(k,p)$, i.e. $\Gamma=A^{-1}$. On the
other hand, the linearized version of the gap equation has the form
\begin{equation}
\int dk A(k,p) \Delta(k) = 0,
\end{equation}
in terms of the same operator. Therefore, whenever there is a non-vanishing solution of the
linearized gap equation $A(k,p)$ develops a zero eigenvalue, its inverse becomes singular, and
$\Gamma$ develops a divergence. Therefore the response function $\Pi(0,0)$, which is the just the
integral of $\Gamma$, also develops divergences at the $\beta_c^{(i)}$ whenever the linearized gap
equation has a solution, i.e. at the critical points for the different instabilities. 

\begin{figure}[h]
\begin{center}
\includegraphics[width=8cm]{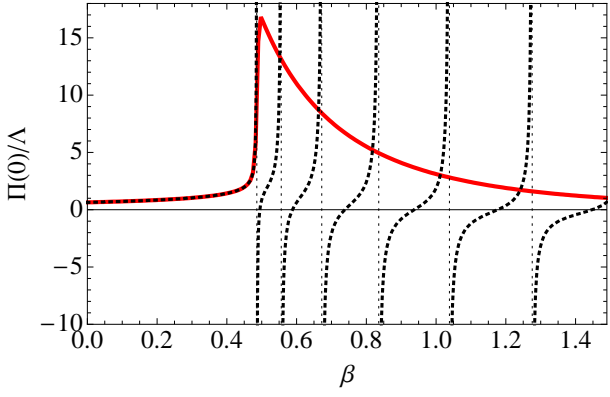}
\caption{Zero momentum response function $\Pi(0,0)$ as a function of $\beta$. When the gap is not
included in the electron propagators, the result presents logarithmically spaced poles (dotted black
line). Asymptotes are shown (thin dotted lines) where poles occur. When self-energy diagrams are
included in the fermion propagators, the generation of a mass prevents the appearance of poles after
the transition (full red line). Note the first pole is approximately located
at the critical coupling obtained from the gap equation. }\label{responsepoles}
\end{center}
\end{figure}

The computation of $\Pi(0,0)$ with $\Delta$ set to zero was carried out in Ref. \onlinecite{WFM10},
where it was proven that an infinite number of logarithmically spaced poles appear for
$\beta>\beta_c$. This result is reproduced in Fig. \ref{responsepoles}. Moreover, the analytical
solution in terms of the model kernel showed that these poles are in fact given also by our Eq.
(\ref{poles}) for the gap equation, as they should. The artificial constraint $\Delta=0$ may
therefore be used to locate the critical couplings for higher order solutions of the gap equation.
Physically, however, as long as we cross $\beta_c$ and the gap is generated, the system always
stays in the lowest energy ground state and no further poles should be observed. This is indeed what
is obtained from the numerical solution of Eq. (\ref{gamma00}) if $\Delta$ is included as computed
from Eq. (\ref{intgap}). In Fig. \ref{responsepoles} we show both cases for comparison. 

\subsection{General computation of $\Gamma$ and $\Pi$}

We now proceed to solve the general equations (\ref{response}) and (\ref{selfc}). We will only
consider $\beta< \beta_c$ for simplicity, i.e., we set $\Delta=0$ for the rest of the article. To
solve this set of equations, it is convenient to decompose $\Gamma_m$ in a basis of 4x4 matrices
with well defined transformation properties under the SU(2) valley symmetry. Defining
$\tilde{M}=\tau_z \sigma_z$, this
basis may be taken as the four matrices
$\tilde{M},\mathcal{I},\alpha^i$ which are scalars under this symmetry, and the matrices
$M_n,T_n,\alpha^i T_n$, each of which transforms like a spin 1/2. With this choice we
express $\Gamma_m$ as
\begin{equation}\label{decomp}
\begin{split}
\Gamma_m &= \tilde{\Gamma}_m \tilde{M} + \tilde{\Gamma}^0_m \mathcal{I} + 
\tilde{\Gamma}^i_m \alpha^i \\
& +\Gamma_{nm} M_n+ \Gamma^{0}_{nm} T_n + \Gamma^{i}_{nm} \alpha^i T_n.
\end{split}
\end{equation}
The equations are further
simplified when $\Gamma^{i}_{nm}$ is expressed in terms of its longitudinal and transverse parts
\begin{align}\label{LT}
\Gamma^L_{nm} = \hat q \cdot \vec{\Gamma}_{nm}, &  & 
\Gamma^T_{nm} = \hat q \times \vec{\Gamma}_{nm},
\end{align}
where $\hat q = \vec q / q$. A similar relation applies for $\tilde{\Gamma}^i_{nm}$. With the
identities
\begin{align}
\vec k \cdot \vec \Gamma_{nm} &= \vec k \cdot \hat q \; \Gamma^L_{nm} - \vec k \times \hat q \;
\Gamma^T_{nm}, \\
\vec k \times \vec \Gamma_{nm} &=\vec k \cdot \hat q \; \Gamma^T_{nm} + \vec k \times \hat q \;
\Gamma^L_{nm},
\end{align}
substituting Eq. (\ref{decomp}) into Eq. (\ref{response}), and
performing the trace, we obtain
\begin{align}\label{response2}
\Pi_{nm}(q) = i  \int & \frac{d^3p}{(2\pi)^3} \;  \frac{8}{D}\left[ f_{11}\Gamma_{nm} + f_{12}
\Gamma^T_{nm}  \right. \nonumber \\ 
& \left. +  \vec p \times \vec q (f_{13}\Gamma^L_{nm} + f_{14}\Gamma^0_{nm}) \right], 
\end{align}
 where we have defined the denominator 
\begin{equation}
D(p,q) = [p_0^2-\vec{p}^2+i\epsilon][(p_0+q_0)^2-(\vec p + \vec
q)^2+i\epsilon],
\end{equation} and where all $f_{ij}(\vec p , \vec q)$ (specified below) are even functions under
the reversal of the
relative angle
$\theta_{\vec p, \vec q} = \theta_p - \theta_q$. Because of the decomposition in Eq. 
(\ref{decomp}), the scalar parts decouple completely and are not needed. We can then obtain
equations for the relevant components of $\Gamma_m$ by multiplying Eq. (\ref{selfc}) by the
corresponding basis matrices and taking the trace. One then obtains
\begin{align}\label{gammanm}
\Gamma_{nm} & = \delta_{nm} -  i\beta \int \frac{d^3 k}{(2\pi)^2}  \frac{1}{D}\frac{1}{|\vec p-\vec
k|}\left[ f_{11}\Gamma_{nm} + f_{12} \Gamma^T_{nm}  \right. \nonumber \\ 
& \left. +  \vec k \times \vec q (f_{13}\Gamma^L_{nm}- f_{14}\Gamma^0_{nm}) \right], \\
\label{gammat}
\Gamma^T_{nm} & = -i\beta \int  \frac{d^3 k}{(2\pi)^2} \frac{1}{D}\frac{1}{|\vec p-\vec
k|}\left[f_{21}\Gamma_{nm}+ f_{22} \Gamma^T_{nm} \right. \nonumber \\ 
& \left. + \vec k \times \vec q \left(f_{23} \Gamma^L_{nm} + f_{24} \Gamma^0_{nm})\right)
\right].
\end{align} 
$\Gamma^L_{mn}$ and $\Gamma^0_{mn}$ satisfy similar
equations, but are not needed in what follows. We now perform a
circular harmonic expansion
\begin{equation}
\Gamma^{(n_p,n_q)} = \int \frac{d\theta_p}{2\pi}e^{i n_p\theta_p}\frac{d\theta_q}{2\pi}e^{i
n_q\theta_q}\Gamma(p, q), 
\end{equation}
and retain only the first order contribution. Terms containing $\vec k \times \vec q$ are odd and
vanish. Thus, $\Gamma^L_n$ and $\Gamma^0_n$ completely decouple to first order.
As anticipated, from the structure of Eqs. (\ref{response2}), (\ref{gammanm}) and (\ref{gammat}) it
can be seen that in fact $\Pi_{nm} = \delta_{nm} \Pi$. With this simplification the relevant
components of $f_{ij}$ are
\begin{align}
f_{11} &=-k_0(k_0+q_0)+ \vec k (\vec k + \vec q\,), \\
f_{12} &=f_{21} =i(\frac{q_0 \vec k \vec q}{q}-k_0q), \\
f_{22} &= \frac{2(\vec q \times \vec k)^2}{q^2}+k_0(k_0+q_0)- \vec k (\vec k + \vec q\,).
\end{align}
Defining
\begin{equation}
K_{ij}^{(n)} = \frac{i}{\pi} \int \frac{d\theta_p}{2\pi} e^{\theta_p n}\int dk_0 k
\frac{f_{ij}}{D},\label{kker}
\end{equation}
the self-consistent equations to first order in the circular harmonic expansion finally read
\begin{align}
\Gamma^{(0,0)} = 1+\frac{\beta}{2p}\int dk C^{(0)} (K_{11}^{(0)} \Gamma^{(0,0)} +
K_{12}^{(0)}\Gamma_T^{(0,0)} ), \\
\Gamma_T^{(0,0)} = -\frac{\beta}{2p} \int dk C^{(0)} (K_{21}^{(0)}\Gamma^{(0,0)}+
K_{22}^{(0)}\Gamma_T^{(0,0)}
),\end{align}
where the Coulomb kernel $C^{(n)}$ was defined in Eq. (\ref{coulombker}). The mixing Kernel
$K_{12}$, as well as the higher order harmonics of the kernel $K_{11}$ can be shown to be
small and may be neglected. In this
case, the final equations determining the response function, spelling momenta explicitly, are
\begin{align} 
\Gamma^{(0,0)}(p,q) &= 1+\frac{\beta}{2p}\int dk C^{(0)}(k/p) K_{11}^{(0)}(k,q) \Gamma^{(0,0)}(k,q),
\label{finalg} \\
\Pi(q) &= \frac{2}{\pi} \int d p K_{11}^{(0)}(p,q) \Gamma^{(0,0)}(p,q). \label{finalm}
\end{align}
When the external $q<q_0$, all $K_{ij}$ develop an imaginary part for
$(q_0-q)/2 < k < (q_0+q)/2$. Note that when $\omega=q=0$, $K_{11}^{(0)}(k,q)=1$ and we recover Eq.
(\ref{gamma00}) for $\Delta=0$. 

\subsection{Analytic solution for the static vertex and response function for $\beta < \beta_c$}

In this section, we show how equations (\ref{finalg}) and (\ref{finalm}) can be solved analytically
in the static limit $q_0=0$, if one assumes simplified versions for the kernels in the spirit of the
previous section. This type of solution is also related to the one employed in the Lorentz invariant
case in QED3 in the computation of the propagator of fermion-antifermion composites \cite{GHR01}
(the analog of excitons in our case). The inclusion of an infrared cutoff in this approximation
makes it excessively complicated, so we will set $\mu=0$ for this section. 

\begin{figure}[h]
\begin{center}
\includegraphics[width=8cm]{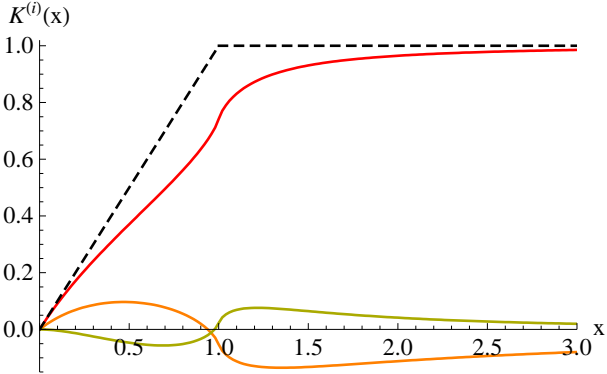}
\caption{Plots of the first three circular harmonics of the kernel $K^{(i)}_{11}(k/q)$, Eq. 
(\ref{kw0q}). The corresponding simplified kernel (\ref{simpker}) is shown for comparison (dashed
line).}\label{kernelplots2}
\end{center}
\end{figure}

The explicit expression for the static kernel in Eq. (\ref{finalg}) is
\begin{align}
\label{kw0q}
K^{(n)}_{11}(k/q) &=  \int \frac{d \theta_k}{2\pi} \frac{e^{i n\theta_k} k}{\vec q^2 + 2 \vec q \vec
k}\left(\frac{\vec k \vec q}{k}+\frac{\vec q (\vec
k +\vec q)}{|\vec k +\vec q|}\right).
\end{align}
Note we can write $K^{(0)}_{11}(k,q)= K^{(0)}_{11}(k/q)$ when $q_0=0$. We will use the following
simplified version
\begin{align}\label{simpker}
K^{(0)}_{11}(x) = x \theta(1-x) + \theta(x-1),
\end{align}
which is compared with the actual kernel and its higher order harmonics that are neglected in Fig.
\ref{kernelplots2}. 

We define $\Gamma \equiv \Gamma^{(0,0)}$ for convenience. Its integral equation in terms of the
simplified kernels is 
\begin{align} 
\Gamma(p,q) =& 1 + \frac{\beta}{2p} \int_0^q dk\frac{k}{q}C^{(0)}(k/p)\Gamma(k,q) \nonumber \\ 
+& \frac{\beta}{2p} \int_q^{\Lambda} dk C^{(0)}(k/p) \Gamma(k,q), \label{simpgamma}
\end{align}
which can be seen to reduce to the usual equation when $q \rightarrow 0$. The
advantage of the simplified kernel $K^{(0)}_{11}$, as we now show, is that we can separate
$\Gamma$ into its $q>0$ and $q<0$ parts 
\begin{equation}
\Gamma(k,q) = \Gamma^{<}(k,q)\theta(q-k) + \Gamma^{>}(k,q)\theta(k-q),
\end{equation}
and obtain two equations that can be solved separately and then matched. The equation for
$\Gamma^<(p,q)$ is simply Eq. (\ref{simpgamma}) when $p<q$ 
\begin{align} 
\Gamma^<(p,q) &= 1 + \frac{\beta}{2}\int_0^q dk\frac{k}{q}
\left(\frac{\theta(p-k)}{p}+\frac{\theta(k-p)}{k}\right) \Gamma^<(k,q) \nonumber\\
+& \frac{\beta}{2}\int_q^{\Lambda}dk\frac{1}{k}\Gamma^>(k,q), 
\end{align}
while for $p>q$ we have
\begin{align} 
\Gamma^>(p,q) &= 1+\frac{\beta}{2}\int_0^q dk \frac{k}{pq}\Gamma^<(k,q) \nonumber \\ 
+& \frac{\beta}{2}\int_q^{\Lambda} dk\left(\frac{\theta(p-k)}{p}+\frac{\theta(k-p)}{k}\right)
\Gamma^>(k,q).
\end{align}
If we define
\begin{align}\label{alpha1}
\alpha_1(q) = \int_0^q dk \frac{k}{q^2} \Gamma^<(k,q), \\
\alpha_2(q) = \int_q^{\Lambda} dk \frac{1}{k}\Gamma^>(k,q),\label{alpha2}
\end{align}
we can write the two equations as
\begin{align} 
\Gamma^<&(p,q) = 1+\frac{\beta}{2}\alpha_2(q) \nonumber \\
+& \frac{\beta}{2}\int_0^q dk\frac{k}{q}
\left(\frac{\theta(p-k)}{p}+\frac{\theta(k-p)}{k}\right) \Gamma^<(k,q), \label{less}\\
\Gamma^>&(p,q) = 1+\frac{\beta}{2}\frac{q}{p} \alpha_1(q) \nonumber \\ 
+& \frac{\beta}{2}\int_q^{\Lambda}
dk\left(\frac{\theta(p-k)}{p}+\frac{\theta(k-p)}{k}\right) \Gamma^>(k,q). \label{more}
\end{align}
Taking two derivatives with respect to $p$ we obtain
\begin{align}
p^2 \frac{\partial^2\Gamma^<}{\partial p^2} + 2p \frac{\partial\Gamma^<}{\partial p}  +
\frac{\beta}{2} \frac{p}{q}\Gamma^<=0, \\
p^2 \frac{\partial^2\Gamma^>}{\partial p^2} + 2p \frac{\partial\Gamma^>}{\partial p}  +
\frac{\beta}{2}\Gamma^>=0 .
\end{align}
Thus we have obtained two independent equations for $\Gamma^<$ and $\Gamma^>$. These equations are
only coupled through the boundary conditions, which can be obtained from the first derivatives of
equations (\ref{less}) and (\ref{more}). For $\Gamma^<$ these are
\begin{align}
 \left.  \left( p \frac{\partial \Gamma^<}{\partial p} + \Gamma^< -1-\frac{\beta}{2}\alpha_2
\right)
\right|_{q} = 0, \\
\left.  p^2 \frac{\partial \Gamma^<}{\partial p}  \right|_{0} = 0,
\end{align}
while for $\Gamma^>$ they are
\begin{align}
\left. \left( p \frac{\partial \Gamma^>}{\partial p} + \Gamma^> -1 \right) \right|_{\Lambda} = 0, \\
\left.  p \frac{\partial \Gamma^>}{\partial p} +\frac{\beta}{2}\alpha_1 \right|_{q} = 0.
\end{align}

The differential equations have straightforward solutions. We note the equation for $\Gamma^>$ is
again the same as the one obtained in the gap equation, and the one obtained in Ref.
\onlinecite{WFM10}, with $q$ playing the role of the infrared cutoff. The equation for $\Gamma^<$ is
a Bessel-type equation. Their solutions are
\begin{align}
\Gamma^> &= A_+ \tilde{p}^{\eta_+} + A_- \tilde{p}^{\eta_-}, \\
\Gamma^< &= \left(\frac{p}{q}\right)^{-1/2} \left(c_1 J_1\left(\sqrt{2\beta p/q}\right) + c_2
Y_1\left(\sqrt{2\beta
p/q}\right)\right).
\end{align}
Applying the boundary conditions we get the solutions
\begin{align}
\Gamma^> = 
\frac{\left( \tilde{q}^{\eta_-}-\alpha_1(1+\eta_-)^2\right)(1+\eta_+) \tilde{p}^{\eta_+}}{
 \tilde{q}^{\eta_-}(1+\eta_+)^2- \tilde{q}^{\eta_+}(1+\eta_-)^2} 
+ (\eta_+ \leftrightarrow \eta_-),
\end{align}
\begin{align}
\Gamma^< = \frac{(1+\beta \alpha_2/2)J_1(\sqrt{2\beta p
/q})}{J_1(\sqrt{2\beta})-\sqrt{\beta/2}J_2(\sqrt{2\beta})}
\left(\frac{p}{q}\right)^{-1/2} .
\end{align}
These solutions still depend on $\alpha_{1,2}$. Plugging them into Eqs. (\ref{alpha1}) and
(\ref{alpha2}) we obtain a linear
system of equations for $\alpha_{1,2}$, whose solutions are
\begin{align}
\alpha_1 &= \frac{(2/\beta-1)\left((1+\eta_+)^2-\tilde{q}^\gamma (1+\eta_-)^2\right)+ \gamma
\tilde{q}^{\eta_+}}{\left((1+\eta_+)^2-\tilde{q}^\gamma (1+\eta_-)^2 \right)/\phi(\beta) +
\beta^2/4(\tilde{q}^\gamma-1)},\\
\alpha_2 &= \phi(\beta) \alpha_1-\frac{2}{\beta},
\end{align}
with
\begin{equation}
\phi(\beta) = \sqrt{2/\beta} \frac{J_1(\sqrt{2\beta})}{J_2(\sqrt{2\beta})}-1.
\end{equation}
The response function can be finally obtained as
\begin{equation}
\Pi(q) = \frac{1}{\pi} \int_0^q dp \frac{p}{q} \Gamma^<(p,q) +\frac{1}{\pi} \int_q^{\Lambda} dp
\Gamma^>(p,q).
\end{equation}
Evaluating the integral, plugging the values of $\alpha_{1,2}$ and in the limit $q<<\Lambda$, we
finally obtain
\begin{equation}
\Pi(q) = \frac{\Lambda}{\pi} \frac{1}{(1+\eta_+)^2} \left[ 1 + \left(\gamma^2 /\phi(\beta)
- 1\right)
\tilde{q}^{\gamma}\right].
\end{equation}
This result reproduces the power law behavior of $\Pi(\omega=0,q)$ found numerically in Ref.
\onlinecite{WFM10}, and shows analytically that the exponent is in fact given by $\eta_0 = \gamma =
\sqrt{1-\beta/\beta_c}$ with $\beta_c =1/2$. This is the excitonic transition again, in limit
$\mu \rightarrow 0$: when $\beta= \beta_c$ the response function becomes singular. This
analytical expression provides a simple, compact expression for observables that couple to the
static response function in the ladder approximation.  

\subsection{Dynamic response}

\begin{figure}[h]
\begin{center}
\includegraphics[width=8cm]{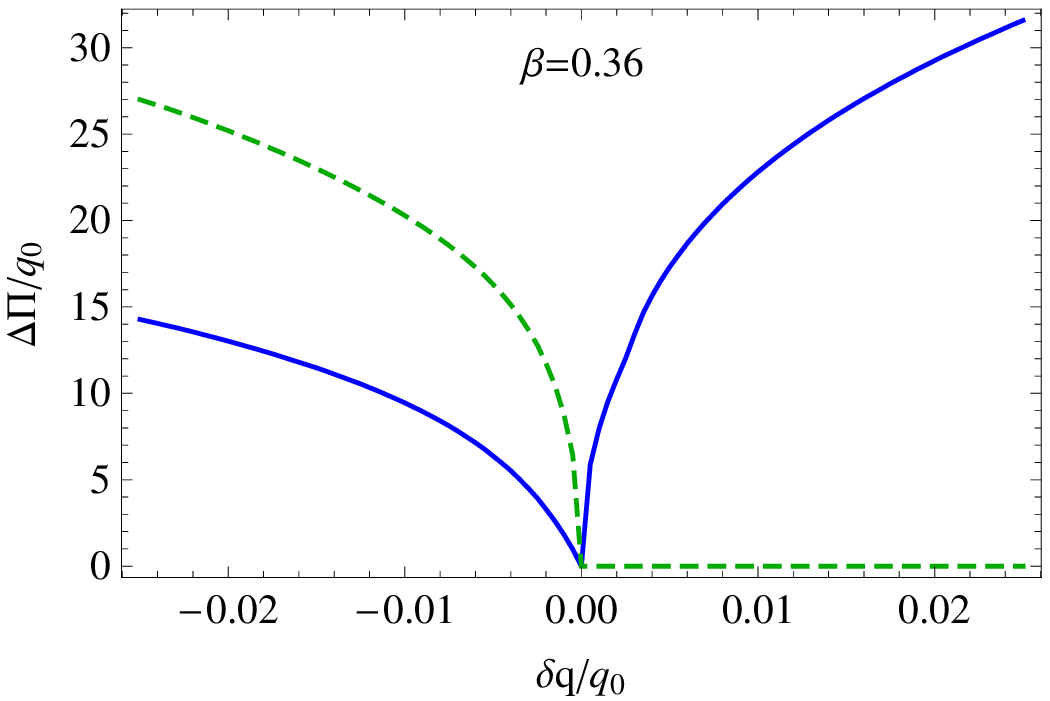}
\includegraphics[width=8cm]{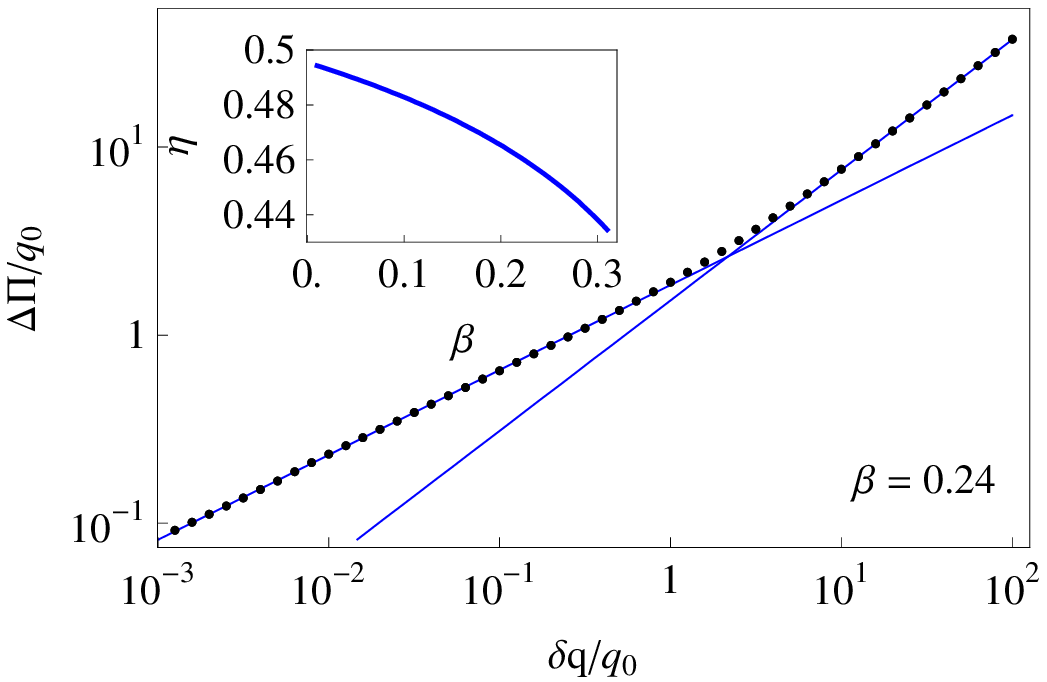}
\caption{Response function $\Delta \Pi(q_0,q_0 + \delta q)$. a) $\Delta \Pi$
for $|\delta q| << q_0$ and $\beta = 0.36$, real part (full line) and imaginary part (dashed
line). Inset: the Kekulé phonon displacements. b) Logarithmic plot of $\Delta \Pi$ for $\beta =
0.24$ and $\delta q > 0$ (dotted line). The full lines are linear fits with $\eta=0.45$ for $\delta
q << q_0 $ and $\eta_0=0.69$ for $\delta q >> q_0$.}\label{powerlaws}
\end{center}
\end{figure}

The kernels in the general frequency dependent response are too complicated for an
analytic solution. Therefore, we now solve Eq. (\ref{finalg}) numerically by discretizing the
momentum $k$ on a logarithmic mesh and solving the corresponding matrix equation by Gaussian
elimination. The integration of Eq. (\ref{finalm}) is straightforward. The results of this procedure
are shown in Fig. \ref{powerlaws}. The self-energy is
represented as the
difference $\Delta \Pi = \Pi(q_0,q_0 + \delta q)-\Pi(q_0,q_0)$ with $\delta q = q_0 - q$ for
convenience. 

The main result of the inclusion of finite frequency is that the response is strongly modified at
$q$ close to $q_0$, but remains essentially the static result once $q>> q_0$. Fig.
\ref{powerlaws}(a) displays the real and imaginary parts of $\Delta \Pi$ for $|\delta q| <<
q_0$. We observe a cusp at $\delta q=0$ in the real part, and a finite imaginary part for $\delta
q<0$. Log plots of both sides of the real part and and the imaginary part reveal power
laws as $\delta q \rightarrow 0$. A Kramers-Kronig analysis for $|\delta q| <<
q_0$ shows that this is only consistent if $\Delta \Pi \propto (\delta q)^{\eta}$, i.e. the
exponents are all the same \footnote{Similar behavior is observed for Kohn anomalies in some one
dimensional systems, see A. Luther and I. Peschel, Phys. Rev. B \textbf{9}, 2911 (1974).}. Fig
\ref{powerlaws}(b) shows a log plot for $\delta q>0$
where power law behavior is evident for $\delta q << q_0$. We also observe that $\Delta \Pi$
crosses over to a different power law for $\delta q >> q_0$, which we identify as the static result
$q^{\eta_0}$ \cite{WFM10}, obtained analytically in the previous section. The inset of Fig
\ref{powerlaws}(b) shows that $\eta$ is $\beta$-dependent, and that it tends to the non-interacting
result in Eq. (\ref{nonint}) as $\beta \rightarrow 0$. In summary, the main features of the
response function are a cusp $(q-q_0)^{\eta}$ around $q_0$, and a crossover to the static power
$q^{\eta_0}$ for $q>>q_0$.

\section{Experimental signatures of power law behavior}\label{sec4}

In the previous section we found that the mass susceptibilities have characteristic power law
behavior with $\beta$ dependent exponents. We may now ask what are the experimental consequences of
this. In general, these correlations can be observed in a linear response-type experiment, with a
suitable probe that couples to electrons in the form of a mass. This is however difficult, as the
usual experiments rather couple to the electron charge or current. To find a probe
that couples to the masses we need to refer to their microscopic origins: A Kekulé
distortion for $M_1,M_2$, or a sublattice antisymmetric potential for $M_3$. 
 
A first proposal to measure the $\Pi_{33}$ correlator was put forward in Ref. \onlinecite{WFM10}
which involved placing a Coulomb impurity asymmetrically with respect to the sublattices, and
measuring the sublattice charge difference with a STM tip. This measurement is difficult to perform,
as it requires one to resolve the lattice structure in detail. A different possibility that we now
discuss is to probe particular phonons that couple to electrons with a mass vertex. The
self-energy of this type of phonons is precisely given by the mass susceptibility, which then
becomes observable through the dispersion and lifetime of the phonon. These can be measured with
current experimental techniques discussed below. 

\begin{figure}[h]
\begin{center}
\includegraphics[width=8.2cm]{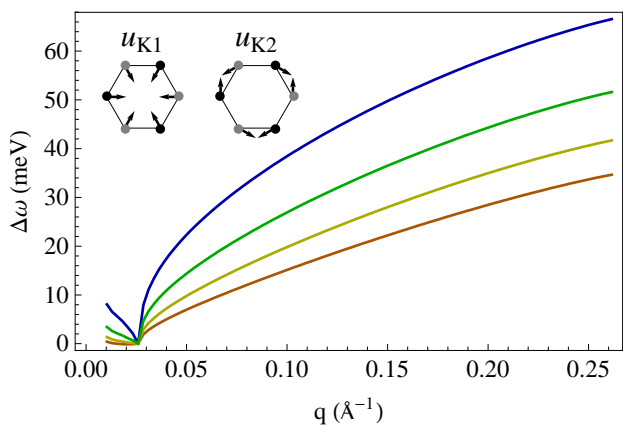}
\caption{$A_1$ phonon dispersion relation $\Delta \omega(q)$ measured from the K point for $\beta =
0,0.1,0.2,0.3$, with higher curves corresponding to higher values of $\beta$. Note that
$\omega(q_K)$, which depends on $\beta$, has been subtracted from each curve for an easier
comparison. Inset: the Kekulé phonon displacements.}\label{phononfig}
\end{center}
\end{figure}

The phonon spectrum of the honeycomb lattice consists
of six phonon branches, four in-plane and two out-of-plane. Each of these phonons may couple to
electrons near either Dirac point if it has momentum close to zero (a $\Gamma$ point or zone
center phonon), which scatters electrons within each valley, or if it has momentum close to $K$ or
$K'$ points (a zone boundary phonon), in which case it produces intervalley scattering. The
strength of the electron-phonon coupling (EPC), however, depends on how the particular 
displacement pattern of that phonon modifies the hopping integrals between atoms. Two modes
have displacements that produce a significant EPC, and both of them are in-plane phonons. The first
of these is the phonon branch of highest energy at the $\Gamma$ point, the $E_2$ phonon. The second
is the $A_1$ branch at the $K$ and $K'$ points (also the highest branch). This is a
lattice distortion with a supercell of six atoms, whose displacement pattern is obtained by taking
linear combinations of the displacements at $K$ and $K'$, and is shown in the inset of Fig.
\ref{phononfig}. These two combinations couple to electrons exactly in the same way as the two
components of the Kekulé distortion, i.e. they couple with the mass matrices $M_1$ and $M_2$
\begin{equation}
H_{e-ph,K} =  F_K \int d^2 r \psi^{\dagger} (M_1 u_{K1} + M_2 u_{K2} )\psi,\label{epv}
\end{equation}
with $F_K = 3 \partial t /\partial a$. For this reason this phonon is also known as the
Kekulé phonon \cite{SA08}. This phonon is therefore suitable to measure the $\Pi_{11}=\Pi_{22}$
correlators. 

It has also been noted that in the presence of broken $z\rightarrow -z$ symmetry,
induced for example by a substrate or a perpendicular electric field, there is another mechanism
that produces an EPC with out-of-plane phonons \cite{FL07,L11}. This is simply that atoms
displaced to different positions see a different potential, and is not related to changes in
hopping integrals. In particular, for the $ZO$ (out-of-plane, optical) phonon at the $\Gamma$ point,
the sublattices A and B move in opposite directions, and there is a linear coupling of the form
\begin{equation}
H_{e-ph,ZO} = F_{ZO} \int d^2 r \psi^{\dagger} M_3 u_{ZO} \psi,\label{epzo}
\end{equation}
with $F_{ZO} \propto E_z$. Therefore, in the presence of an electric field, the $ZO$ phonon can be
used to measure $\Pi_{33}$. 
The Hamiltonian of either phonon may be expressed as 
\begin{align}
H = \sum_i \int \frac{d^2q}{(2\pi)^2} \omega_{i} b^{\dagger}_{i,q} b_{i,q},
\end{align}
with creation and destruction operators defined by
\begin{align}
 u_i =  \sqrt{\frac{A_c}{4 \omega_i M}} \int \frac{d^2q}{(2\pi)^2} (b_{i,q}e^{i \vec q \vec r} +
b^{\dagger}_{i,q}e^{-i \vec q \vec r}),
\end{align}
where $i=K1,K2,ZO$, $\omega_K \approx 0.17$ eV, $\omega_{ZO} \approx 0.1 $ eV, $A_c$ is the unit
cell area, and the dispersion of the phonons is neglected to a first approximation. A dimensionless
EPC can be defined as $\lambda_K =F_i^2 A_c/(2 M \omega_i v_F^2)$. In the case of the $A_1$ phonon
it is estimated to be in the range $\lambda_{K} \approx 0.03-0.1$ \cite{BA08,BPF09}. We will now
discuss the only the $A_1$ phonon, as the $ZO$ case in an electric field has the same behavior.

The phonon propagator can be obtained in terms of the self-energy as
\begin{equation}\label{phonon}
G_{ph}(\omega,q) = \frac{2\omega_K}{\omega^2 - \omega_K^2 -2\omega_K\Sigma(\omega,q)}.
\end{equation}
and it follows directly from the form of the electron-phonon vertex Eq. (\ref{epv}) that the phonon
self-energy is directly related to the mass susceptibility Eq. (\ref{response}) via $\Sigma =
\frac{\lambda_K}{2} \Pi$. The dispersion relation for the
phonon can be obtained by solving for the pole in Eq. (\ref{phonon}) for small $\lambda_K$,
so that the dispersion relation is corrected to
\begin{equation}
\omega(q) \approx \omega_K + \frac{\lambda_K}{2}\Pi(\omega_K,q). 
\end{equation}
For $\beta=0$ this is \cite{PLM04}
\begin{equation}\label{nonint} 
\omega(q) = \omega_K + \frac{\lambda_K}{4} (v_F^2 q^2 -\omega_K^2)^{1/2},
\end{equation}
which has a square root singularity
at $q_K$ for $q>q_K$. For $q<q_K$ the self-energy is purely imaginary, and a finite lifetime is
obtained. The Kohn anomaly is conventionally associated with a linear cusp in the dispersion, which
is obtained only asymptotically for $q>>q_K$; the full dynamical self-energy should be used in
general. Note that $q_K$ is
approximately 2\% of the $\Gamma-K$ distance in the Brillouin
zone. The necessity of employing the dynamical self-energy has been emphasized before
\cite{LM06,CG07,THD08}, in particular in the doped case where
the static approximation produces poor agreement with experiments \cite{PLC07}. Note also that a
different Dirac fermion induced Kohn anomaly has been recently observed in the surface
of topological insulator Bi$_2$Se$_3$ \cite{ZSS11}.

In the presence of electron-electron interactions, $\Sigma$ is modified as described in the previous
section and in particular acquires power law behavior.  In Fig. \ref{phononfig} we plot the
phonon dispersion relation obtained from it for different values of $\beta$. This is given in terms
of the self-energy evaluated at the phonon frequency $\omega_K$. To ease the comparison at different
values of $\beta$, we also represent the difference  
\begin{equation}
\Delta \omega(q) = \omega(q) - \omega(q_K) = \frac{\lambda_K}{2} \left( \Pi(\omega_K,q) -
\Pi(\omega_K,q_K) \right),
\end{equation}
where we have recovered physical units with $\hbar v_F = 6.5$ eV$\AA$. The values of the parameters
used are $\lambda_K=0.1$ and $\Lambda = 1.7 eV$. The dispersion follows the
static power law $q^{\eta_0(\beta)}$ for $q>>q_K$, and the cusp turns into $q^{\eta(\beta)}$ as
discussed above. The modification of the Kohn anomaly due to interactions is rather dramatic and
should be observable.

From the experimental point of view, there are several techniques available for the measurement of
the $A_1$ phonon dispersion, and each one has its own potential difficulties. One method is to
employ Electron Energy Loss Spectroscopy (EELS), which has already been used to map K-point phonon
dispersions of graphene on different substrates where it behaves as
quasi-freestanding\cite{YTI05,PMF11}, such as Pt. (The absence of hybridization with the
substrate is important as it would strongly change the electron band structure and the Kohn anomaly
\cite{AW10}). The problem with metallic substrates is that screening will turn
the long range Coulomb interaction between graphene electrons into a short-ranged one, or may even
suppress it completely. Thus a power law anomaly is not expected, which is consistent with the
linear
one observed in Ref. \onlinecite{PMF11}. An insulating substrate would be required to observe this
effect. 

A more indirect experiment (with insulating substrate) is to track
the dependence of the 2D Raman peak with incoming laser energy. This method has been used
\cite{MSM07} to measure the dispersion of the $A_1$ phonon, but both the amount of data it yields
and the range of momenta it covers is limited and not very close to the $K$ point. Finally, X-rays
are a usual tool to measure phonon dispersions in 3D crystals, and while it is probably challenging
to obtain enough intensity from a single sheet of graphene, experiments in graphite
\cite{MET04,GSB09} might be used to deduce the phonon dispersion. This approach is not
straightforward because the electronic structure of graphite is different from graphene, and this
must be taken into account. Nevertheless, it is encouraging to observe that precision measurements
show an $A_1$ phonon dispersion that is not linear \cite{GSB09}.

\section{Conclusions and discussion}\label{sec5}

Understanding the role of interactions in graphene is a challenging problem whose solution is far
from complete. One of the main purposes of this work is to bring greater attention to a new
potential way of observing interaction effects: the measurement of the mass susceptibility, which,
as we have shown, displays power law behavior with $\beta$-dependent exponents. Apart for unveiling
a novel signature of interactions, the identification of these power laws could represent an
alternative measurement of graphene's fine structure constant $\beta$. 

Moreover, the observation of the static power law would also help in understanding the
problem of the excitonic transition in graphene. With the experimental evidence gathered so
far, there seems to be no indication of the presence of an interaction induced gap, even in high
quality samples \cite{EGM11}. Theoretically, however, it appears that suspended graphene should be,
if not in the gapped phase, at least close to the transition. Since, as we have shown, the static
exponent of the mass-mass response goes to zero at the critical $\beta_c$, a measurement of the
exponent would indicate how close we are to the achievement of the long-sought gap.

Note also that other potential measurements of these power laws could come from spin related
experiments, which we have not discussed. The inclusion of spin in the picture is straightforward:
in this case, the symmetry of the Hamiltonian enlarged to SU(4), and the excitonic masses generalize
to singlet and triplet versions. Because of the same symmetry argument relating CDW and Kekulé
masses, any of these correlators would have the same power law behavior, so the observation of these
particular spin susceptibilities may represent an alternative route to the experiments we have
discussed in this work.

A final comment concerns the robustness of our result to more refined approximations
schemes than the ladder summation. While other sets of diagrams may modify our quantitative
predictions, it is very unlikely that the non-analytic behavior can be removed in this way. One may
consider, for example, the inclusion of self-energy terms for the electron propagator \cite{GGV94},
which may produce a slow logarithmic dependence of the exponent. This could be taken into
account in a similar way as was done for the gap equation \cite{SSG10}. The effect of static RPA
screening of the Coulomb interaction will in general reduce the effective value of $\beta$ but will
not change the power law itself. Finally we also note that the 1/N approximation does gives power
law behavior for the Kekulé mass correlator \cite{G10,GMP10} (and thus the self-energy) as well.

In summary, we believe that the observation of power law correlations originating in the mass
susceptibility is potentially feasible and would be an important contribution to our understanding
of the problem of interactions in graphene.

\section{Acknowledgments} This work was supported by the NSF through Grant No. DMR1005035,
and by US-Israel Binational Science Foundation (BSF) through Grant No. 2008256.

\bibliography{kohn}

\end{document}